\documentclass[aps,prx,twocolumn,amsmath,amssymb,superscriptaddress,longbibliography]{revtex4-2}
\usepackage[english]{babel}
\usepackage{xcolor}
\usepackage{amssymb}
\usepackage{dcolumn}
\usepackage{bm}
\usepackage{graphicx}
\usepackage{amsmath}
\usepackage{braket}

\allowdisplaybreaks[4]   %Formula span

\usepackage{graphicx}        % standard LaTeX graphics tool

\graphicspath{{pict/}{}}

\usepackage[normalem]{ulem}
\usepackage{dcolumn}%new
\usepackage{bm}
\usepackage[pdfstartview=FitH, CJKbookmarks=true, bookmarksnumbered=true, bookmarksopen=true, colorlinks=true, pdfborder=001, citecolor=blue, linkcolor=blue, urlcolor=blue, linktocpage=true] {hyperref}

\begin{document}

\title{Symmetry-protected Landau-Zener-Stückelberg-Majorana interference and non-adiabatic topological transport of edge states}

\author{Shi Hu}
\email{hush27@gpnu.edu.cn}
\affiliation{School of Optoelectronic Engineering, Guangdong Polytechnic Normal University, Guangzhou 510665, China}

\author{Shihao Li}
\affiliation{School of Optoelectronic Engineering, Guangdong Polytechnic Normal University, Guangzhou 510665, China}

\author{Meiqing Hu}
\affiliation{School of Optoelectronic Engineering, Guangdong Polytechnic Normal University, Guangzhou 510665, China}

\author{Zhoutao Lei}
\email{leizht3@mail2.sysu.edu.cn}
\affiliation{Guangdong Provincial Key Laboratory of Quantum Metrology and Sensing $\&$ School of Physics and Astronomy, Sun Yat-Sen University (Zhuhai Campus), Zhuhai 519082, China}
\date{\today}

\begin{abstract}
We systematically investigate Landau-Zener-Stückelberg-Majorana (LZSM) interference under chiral-mirror-like symmetry and propose its application to non-adiabatic topological transport of edge states.
Protected by this symmetry, complete destructive interference emerges and can be characterized through occupation probability. 
This symmetry-protected LZSM interference enables state transitions to be achieved within remarkably short time scales.
To demonstrate our mechanism, we provide two distinctive two-level systems as examples and survey them in detail.
By tuning evolution speed or increasing holding time, the complete destructive interferences are observed.
Furthermore, we make use of this mechanism for topological edge states of Su-Schrieffer-Heeger (SSH) chain by taking them as an isolated two-level system.
Through carefully designed time sequences, we construct symmetry-protected LZSM interference of topological edge states, enabling non-adiabatic topological transport.
Our work unveils an alternative way to study quantum control, quantum state transfer, and quantum communication via non-adiabatic topological transport.
\end{abstract}
\maketitle

%%%%%%%%%%%%%%%%%%%%%%%%%%%%%%%%%%%%%%%%%%%%%%%%%%%%%%%%%%%%%%%%%%%%%%%%%%%%%%%%%%%%
\section{Introduction}\label{Sec1}
%%%%%%%%%%%%%%%%%%%%%%%%%%%%%%%%%%%%%%%%%%%%%%%%%%%%%%%%%%%%%%%%%%%%%%%%%%%%%%%%%%%%
The Landau-Zener (LZ) transition represents a fundamental quantum dynamical process describing the evolution of a two-level system (TLS) driven across an avoided level crossing~\cite{Landau1932,Zener1932,Stückelberg1932,Majorana1932}. 
When multiple LZ transitions occur in succession, Landau-Zener-Stückelberg-Majorana (LZSM) interference emerges~\cite{ShevchenkoPR2010,IvakhnenkoPR2023}, where the final occupation probability depends not only on individual LZ transition probabilities but also on the phase accumulated during the dynamic process.
Furthermore, if these LZ transitions are structured, certain types of interference results can be generated such as complete remaining in ground state, complete transition to excited state, and $50:50$ quantum beam splitters, with potential application in interferometry and quantum control~\cite{ShevchenkoPR2010,IvakhnenkoPR2023, MarkPRL2007,CaoNC2013,QuintanaPRL2013,TanPRL2014, ForsterPRL2014,BoganPRL2018,CampbellPRX2020,WenPRB2020,LonghiAQT2019,LiuarXiv2024}. 
LZ transition and LZSM interference have been extensively studied across diverse physical platforms, including ultracold molecules~\cite{MarkPRL2007}, quantum dots~\cite{CaoNC2013}, nanostructures~\cite{AverinPRL1995,AverinPRL1999}, Bose-Einstein condensates~\cite{ZenesiniPRL2009}, Rydberg atoms~\cite{ZhangPRL2018}, atomic qubits~\cite{FerrierPRL2013}, graphene~\cite{HiguchiNature2017}, and waveguide systems~\cite{ChenPRL2021,XiePRB2022}.
Moreover, the study has expanded to encompass nonlinear systems~\cite{LiuPRA2002,LiPRA2018,WangNJP2023}, multilevel configurations~\cite{ShytovPRA2004,SunNC2010}, and non-Hermitian frameworks~\cite{WangNJP2023,ShenPRA2019}.
Although many relevant works have been developed for specific systems, it is useful and meaningful to establish a mechanism for discovering and analyzing qualified systems that can generate these intriguing LZSM interferences.

On the other hand, adiabatic topological transport, aiming to transfer particle or quantum state utilizing topological properties of a adiabatical modulated system, has attracted extensive concern in recent years. 
In general, there are two typical kinds of transport: adiabatic pumping via topological edge modes~\cite{YEKrausPRL2012,MVerbinPRB2015} and quantized Thouless pumping via topological bulk bands~\cite{DJThoulessPRB1983,QNiuJPA1984,YKeLPR2016,
YKePRA2017,SHuPRB2019,YKePRR2020}. 
Both of them have been widely applied in many fields of physics, including quantum state transfer~\cite{NLangQI2017,CDlaskaQST2017,FMeiPRA2018,
SLonghiPRB2019,NEPalaiodimopoulosPRA2021,LHuangPRA2022,CWangPRA2022}, quantum interference~\cite{JLTambascoSA2018,SHuPRA2020,SHuPRA2024}, quantum gates~\cite{PBorossPRB2019,MNarozniakPRB2021}, and quantum devices~\cite{RHammerPRB2013,XSWangPRB2017,LQiPRB2021,LQiPRA2023}. 
However, the strict adiabatic requirement often necessitates lengthy evolution times otherwise the transport will be destroyed due to non-adiabatic effects~\cite{LPriviteraPRL2018}.
Thus it is important to find an effective transport approach beyond adiabatic topological transport.

In this work, we demonstrate how chiral-mirror-like symmetry can protect complete destructive LZSM interference.
While chiral-mirror symmetry is a combination of chiral and mirror symmetry~\cite{GeierPRB2018,TrifunovicPRX2019,CoutantAPR2023,LeiPRB2023}, the chiral-mirror-like symmetry reversing the axis of time instead of space [see Eq.~\eqref{Symmetry}].
The flexibility in Hamiltonian structure and symmetry operator choice makes our mechanism widely applicable across various systems.
Notably, complete destructive interference manifests distinct outcomes depending on the relationship between the initial Hamiltonian and symmetry operator.
Based on transfer matrix (TM) method, a unified expression for occupation probability after LZSM interference will be given and then the complete destructive conditions will be obtained.
We validate our mechanism through analysis of two distinct TLSs with different symmetry operators, comparing TM method predictions with numerical evolution results. 
As a practical application, we demonstrate that symmetry-protected LZSM interference can be extended to topological edge states, enabling non-adiabatic topological transport beyond conventional adiabatic approaches.

This paper is organized as follows. 
In Sec.~\ref{Sec2}, we introduce the mechanism, including the constraint of chiral-mirror-like symmetry and the resulting expression of LZSM interference with TM method. 
In Sec.~\ref{Sec3}, we propose two TLSs, involving two band minima or completely flat bands, to elaborate and validate our mechanism. 
And in Sec.~\ref{Sec4}, we apply this mechanism to propose non-adiabatic topological transport of edge states.
Finally, the summary and discussion are included in Sec.~\ref{Sec5}.

%%%%%%%%%%%%%%%%%%%%%%%%%%%%%%%%%%%%%%%%%%%%%%%%%%%%%%%%%%%%%%%%%%%%%
\section{Symmetry-protected Landau–Zener–Stückelberg–Majorana interference}\label{Sec2}
%%%%%%%%%%%%%%%%%%%%%%%%%%%%%%%%%%%%%%%%%%%%%%%%%%%%%%%%%%%%%%%%%%%%%
We begin with a generic TLS described by the time-dependent Hamiltonian ($\hbar=1$)
\begin{eqnarray}\label{Ham}
\hat{H}(t)/J=
d_{0}(t)\hat{\sigma}_{0}+d_{x}(t)\hat{\sigma}_{x}
+d_{y}(t)\hat{\sigma}_{y}+d_{z}(t)\hat{\sigma}_{z},
\end{eqnarray}
and the evolution of wave function is given by
\begin{eqnarray}\label{Psit}
|\Psi(t)\rangle=\hat{T}e^{-i\int_0^t\hat{H}(\tau)d\tau}|\Psi(0)\rangle,
\end{eqnarray}
with the time-ordering operator $\hat{T}$.
Here $J$ ($1/J$) is energy (evolution time) unit, $\hat{\sigma}_{0}$ is identity matrix, $\hat{\sigma}_{x,y,z}$ are Pauli matrices with basis $\{|0\rangle,|1\rangle\}$, and $d_{0,x,y,z}$ are time-dependent real parameters.
We note that when the time-dependent $d_{\alpha=x,y,z}$ satisfies certain conditions, such as associated with a driving along a single axis~\cite{BarnesPRL2012}, and even more generally, along any set of axes of the Bloch sphere~\cite{BarnesPRA2013}, the LZSM interference as well as the complete destructive condition can be determined exactly.
Specifically, the key assumption is that this system preserves chiral-mirror-like symmetry during the evolution, that is
\begin{eqnarray}\label{Symmetry}
\hat{H}(t)&=&-\hat{\sigma}_{r}\hat{H}(T-t)
\hat{\sigma}_{r},\cr\cr
\hat{\sigma}_{r}&=&\sin\theta\cos\varphi\hat{\sigma}_{x}
+\sin\theta\sin\varphi\hat{\sigma}_{y}+\cos\theta\hat{\sigma}_{z}, 
\end{eqnarray}
where $\theta$ and $\varphi$ are azimuth and elevation angles being able to take arbitrary value, respectively.
Thus we can decompose Hamiltonian into  
\begin{eqnarray}\label{Ham2}
\hat{H}(t)/J&=&d_{0}(t)\hat{\sigma}_{0}
+d_{\theta}(t)\hat{\sigma}_{\theta}
+d_{\varphi}(t)\hat{\sigma}_{\varphi}
+d_{r}(t)\hat{\sigma}_{r},\cr\cr
\hat{\sigma}_{\theta}&=&\cos\theta\cos\varphi\hat{\sigma}_{x}
+\cos\theta\sin\varphi\hat{\sigma}_{y}
-\sin\theta\hat{\sigma}_{z},\cr\cr
\hat{\sigma}_{\varphi}&=&-\sin\varphi\hat{\sigma}_{x}
+\cos\varphi\hat{\sigma}_{y},
\end{eqnarray}
where $d_{0,r}(t)=-d_{0,r}(T-t)$ and $d_{\theta,\varphi}(t)=d_{\theta,\varphi}(T-t)$ under the symmetry defined in Eq.~\eqref{Symmetry}.
For the purpose of generality, we do not specify the precise values of $\theta$ and $\varphi$ here.
Under this symmetry, if a non-adiabatic evolution occurs during $[t_i,t_f]$ with $0\leq t_i<t_f\leq T/2$, another one will occur in the range $[T-t_f,T-t_i]$.
Therefore, we can divide the LZSM interference into three stages: two non-adiabatic evolution stages, labeled as I and III, and an adiabatic evolution stage, labeled II, where phase accumulation occurs. 
Stage II may, in fact, be absent.

To study LZSM interference conveniently, we employ the adiabatic basis $\{|E_{-}(t)\rangle,|E_{+}(t)\rangle\}$ with parallel transport gauge~\cite{DXiaoRMP2010}, which is eigenstates of $\hat{H}(t)$: $\hat{H}(t)|E_{\pm}(t)\rangle=E_{\pm}(t)|E_{\pm}(t)\rangle$ and satisfy $\langle E_{\pm}(t)|\partial_t|E_{\pm}(t)\rangle=0$, the adiabatic evolution matrix during stage II is given by:
\begin{eqnarray}\label{U2}
\hat{U}_{\rm{II}}
=\left(
\begin{array}{llllllllll}
e^{-i\phi_{-}}~~~~0~~\cr
~~0~~~~e^{-i\phi_{+}}
\end{array}\right)
=\exp(i\phi_{d}\hat{\sigma}_{z}),
\end{eqnarray}
where $\phi_{d}$ is the dynamical phase accumulated during stage II with
\begin{eqnarray}\label{phid}
\phi_{+}=-\phi_{-},~~~~\phi_{d}=\phi_{+}=\int_{t_f}^{T-t_f} E_{+}(t)dt.
\end{eqnarray}
Here we use the symmetry of the energy spectrum under chiral-mirror-like symmetry, i.e., $E_{\pm}(t)=-E_{\mp}(T-t)$.
Moreover, the chiral-mirror-like symmetry will also enforce a non-trivial relation between two non-adiabatic transition matrices. 
That is the transition matrix for stages I will take the form:
\begin{eqnarray}\label{UI}
\hat{U}_{\rm{I}}=e^{-i\gamma(t_f)\hat{\sigma}_{z}}
\left(
\begin{array}{llllllllll}
\mathcal{R}e^{i\phi_{c}}~~~~
-\mathcal{T}^*e^{i\phi_{c}}  \\
\mathcal{T}e^{-i\phi_{c}}~~~~~
\mathcal{R}^*e^{-i\phi_{c}}
\end{array}\right)
e^{i\gamma(t_i)\hat{\sigma}_{z}},
\end{eqnarray}
and the one for stages III will be
\begin{eqnarray}\label{UIII}
\hat{U}_{\rm{III}}=e^{-i\gamma(t_i)\hat{\sigma}_{z}}
\left(\begin{array}{llllllllll}
\mathcal{R}e^{i\phi_{c}}~~~~
-\mathcal{T}e^{-i\phi_{c}}   \\
\mathcal{T}^*e^{i\phi_{c}}~~~~~
\mathcal{R}^*e^{-i\phi_{c}}
\end{array}\right)
e^{i\gamma(t_f)\hat{\sigma}_{z}},
\end{eqnarray}
where the reflection coefficient $\mathcal{R}$ and transmission coefficient $\mathcal{T}$ can be extracted as $\mathcal{R}=\langle\mathcal{E}_{-}(t_f)|\Psi(t_f)\rangle$ and $\mathcal{T}=\langle\mathcal{E}_{+}(t_f)|\Psi(t_f)\rangle$ when the initial state is set as ground state of $\hat{H}(0)$.
Here $|\mathcal{E}_{\pm}(t_f)\rangle$ is from adiabatic basis with more easily obtained gauge satisfying ${\rm Im}\big[\langle\mathcal{E}_{+}(t_f)|0\rangle\langle0|\mathcal{E}_{-}(t_f)\rangle\big]=0$ and $\det\big[|\mathcal{E}_{-}(t_f)\rangle,|\mathcal{E}_{+}(t_f)\rangle\big]=1$.
In addition, $\gamma(t_{i})$, $\gamma(t_{f})$ and $\phi_{c}\equiv\frac{1}{2}\arg(\cos\theta+i\sin\theta\sin\vartheta)$ with $\vartheta\equiv\arg[d_{\theta}(T/2)-id_{\varphi}(T/2)]$ ($d_{\theta}$ and $d_{\varphi}$ defined in Eq.~\eqref{Ham2}) are the phase factors from gauge transformation (see appendix A).
We note that $\phi_{c}=0$ and $\pi/4$ when $\theta=0$ and $\theta=\pi/2$, respectively, independent of detail of Hamiltonian, while other cases are just dependent on Hamiltonian at half-time point $\hat{H}(T/2)$.

Now, we can obtain the complete evolution matrix of the LZSM interference conveniently by the TM method~\cite{IvakhnenkoPR2023}, where each stage is attributed to the respective matrices, that is
\begin{eqnarray}\label{U}
\hat{U}=\hat{U}_{\rm{III}}\hat{U}_{\rm{II}}\hat{U}_{\rm{I}}
=e^{-i\gamma(t_i)\hat{\sigma}_{z}}\mathcal{\hat{U}}e^{i\gamma(t_i)\hat{\sigma}_{z}},
\end{eqnarray} 
with 
\begin{eqnarray}\label{U1}
\hat{U}=
\left(
\begin{array}{cc}
U_{11}~~~~U_{12}\\
U_{21}~~~~U_{22}
\end{array}\right),
~~~~
\mathcal{\hat{U}}=
\left(
\begin{array}{cc}
\mathcal{U}_{11}~~~~\mathcal{U}_{12}\\
\mathcal{U}_{21}~~~~\mathcal{U}_{22}
\end{array}\right),
\end{eqnarray}
where
\begin{eqnarray}\label{Uelement}
\mathcal{U}_{11}&=&\mathcal{U}^{*}_{22}
=\mathcal{R}^2e^{i(\phi_d+2\phi_{c})}
-\mathcal{T}^2e^{-i(\phi_d+2\phi_{c})},\cr\cr
\mathcal{U}_{21}&=&-\mathcal{U}_{12}
=\mathcal{RT^*}e^{i(\phi_d+2\phi_{c})}
+\mathcal{R^*T}e^{-i(\phi_d+2\phi_{c})}.
\end{eqnarray}
Thereby if we start from the ground state $|E_{-}\rangle$, the probability $\mathcal{P_{-+}}$ of finding the system in its excited state $|E_{+}\rangle$ and remaining in the ground state $\mathcal{P_{--}}$ will be
\begin{eqnarray}\label{P-+}
\mathcal{P_{-+}}&=&|U_{21}|^2=|\mathcal{U}_{21}|^2
=4|\mathcal{R}|^2|\mathcal{T}|^2\cos^2\phi_{\mathcal{L}},\cr\cr
\mathcal{P_{--}}&=&|U_{11}|^2=|\mathcal{U}_{11}|^2=1-\mathcal{P_{-+}},\cr\cr
\phi_{\mathcal{L}}&=&\phi_d+2\phi_{c}+\phi_{r},
\end{eqnarray}
with $\phi_{r}=\arg(\mathcal{RT^*})$ is the relative phase between the reflection coefficient $\mathcal{R}$ and transmission coefficient $\mathcal{T}$.
%
%为了广泛性的兴趣，我们只关注对称性的作用。在特定的情况下，或许可以用逆向工程得到解析的表达式。

As such, the complete destructive LZSM interference corresponding to $\mathcal{P_{--}}=1$ is achieved when $\phi_{\mathcal{L}}=(n+1/2)\pi$, with $n$ being an integer.
After this process, the end state will be ground state of $\hat{H}(T)$ after LZSM interference.
However, this result will be associated with distinct significance associated with different relation between initial Hamiltonian $\hat{H}(0)$ and symmetric operator $\hat{\sigma}_{r}$.
For example, this end state will come back to initial states if they form anticommutation relation $[\hat{H}(0),\hat{\sigma}_{r}]_+=0$.
And more importantly, when they commutate with each other $[\hat{H}(0),\hat{\sigma}_{r}]=0$, indicating $\hat{H}(T)=-\hat{H}(0)$, this process actually is associated with a state transition $|E_{-}(0)\rangle\rightarrow|E_{-}(T)\rangle=|E_{+}(0)\rangle$.
In following of this work, we will mainly focus on the latter situation.

%%%%%%%%%%%%%%%%%%%%%%%%%%%%%%%%%%%%%%%%%%%%%%%%%%%%%%%%%%%%%%%%%%%%%
\section{Results of two-level systems}\label{Sec3}
%%%%%%%%%%%%%%%%%%%%%%%%%%%%%%%%%%%%%%%%%%%%%%%%%%%%%%%%%%%%%%%%%%%%%
\begin{figure*}[!htp]
\includegraphics[width=1\columnwidth]{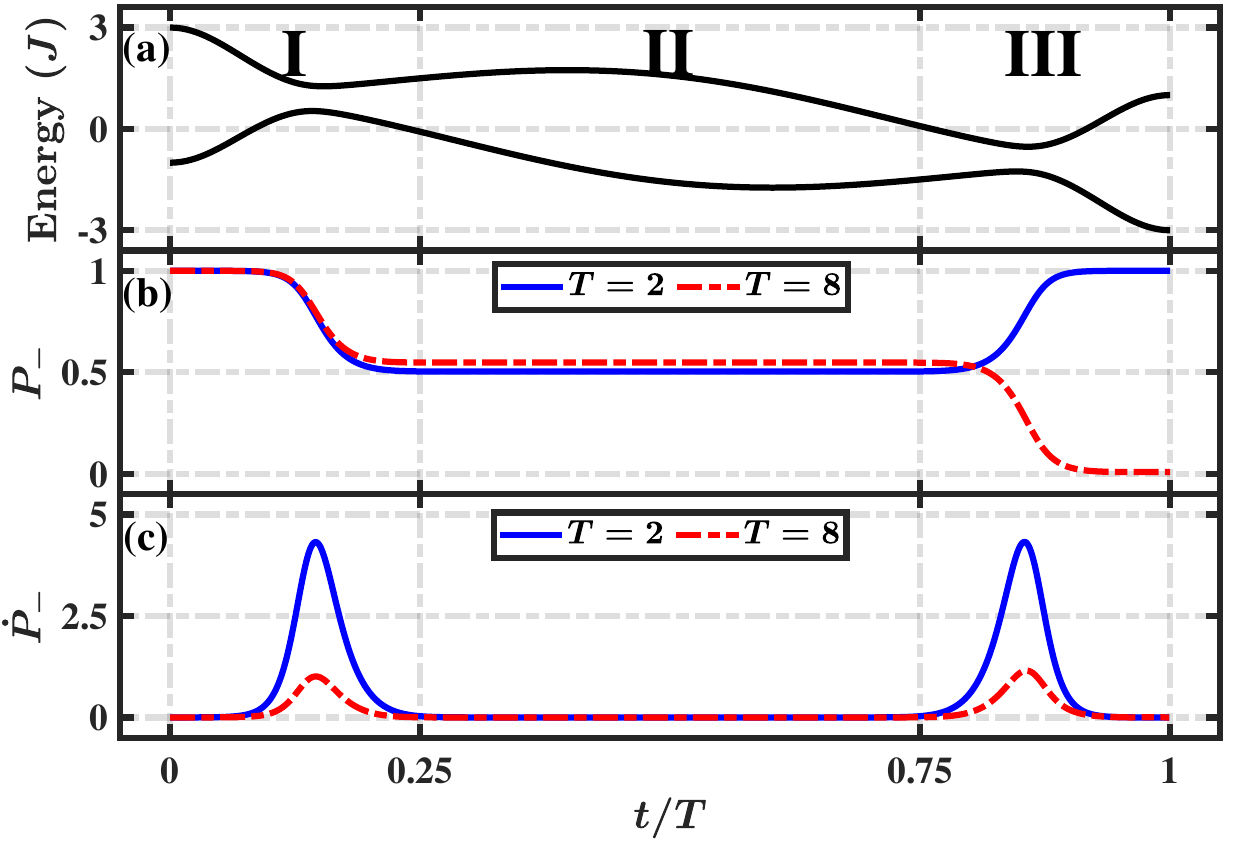}
\includegraphics[width=1\columnwidth]{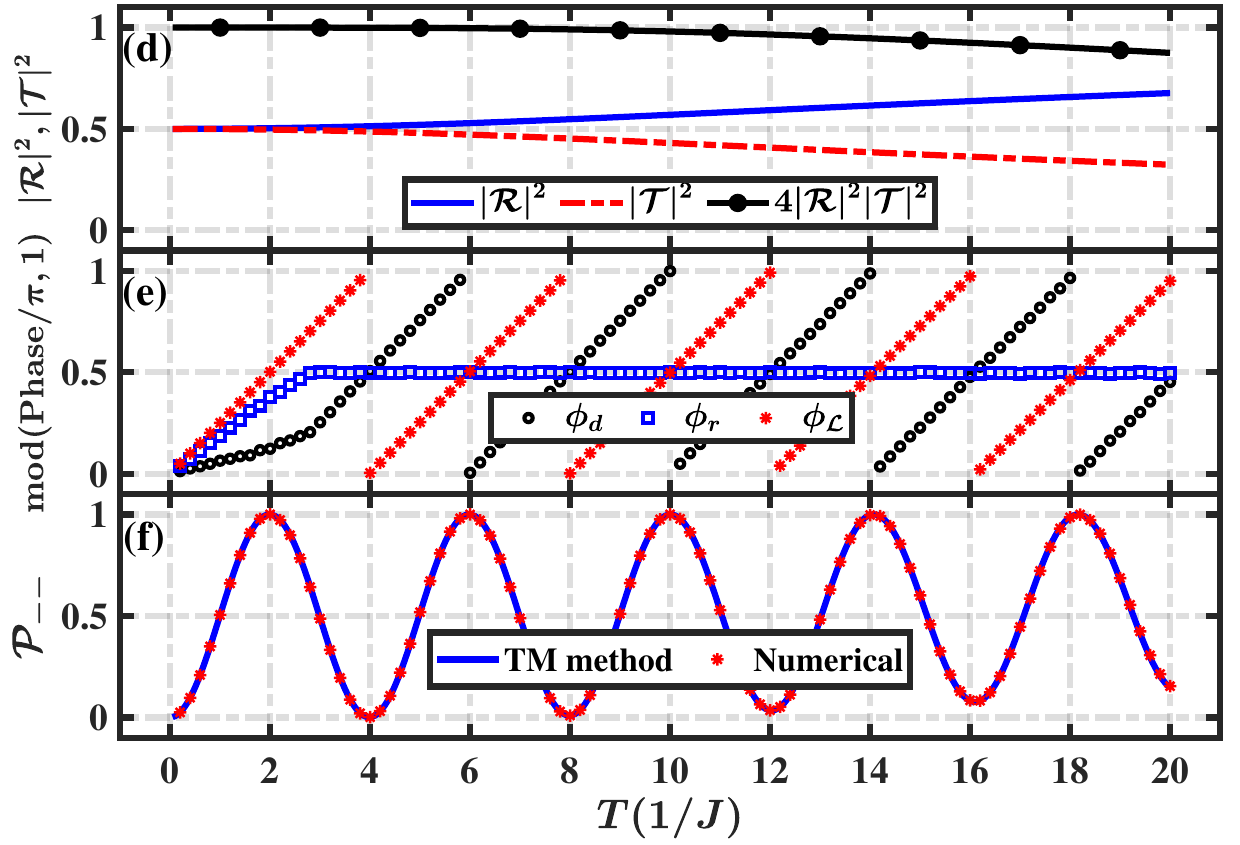}
\caption{\label{Example1}LZSM interference of TLS with two sharp band minima. (a) Energy spectrum for TLS under chiral-mirror-like symmetric Hamiltonian defined by Eqs.~\eqref{Ham} and \eqref{Ham11}. (b) and (c) Occupation probability in ground state $P_{-}$ and its time derivative $\dot{P}_{-}$ versus evolution time $t$ with different total time $T$, where blue solid (red dashed) line corresponds to $T=2$ ($T=8$). (d), (e), and (f) LZSM interference results starting from ground state $|E_{-}\rangle$ versus total evolution time $T$. (d) Reflection and transmission coefficients $|\mathcal{R}|^2$ (blue solid line), $|\mathcal{T}|^2$ (red dashed line) and $4|\mathcal{R}|^2|\mathcal{T}|^2$ (black solid line with dot). (e) Phase factor $\phi_{d}$ (black circle), $\phi_{r}$ (blue square), and  $\phi_{\mathcal{L}}$ (red asterisk). The phase factor from gauge transformation is $\phi_{c}=0$ due to $\theta=0$. (f) Transfer matrix (TM) method solutions (blue solid line) obtained from Eq.~\eqref{P-+} and numerical results (red asterisk) of occupation probability $\mathcal{P_{--}}$. In all panels, parameters are set as $J_{0}=1$, $J_{x}=1.5$, $J_{y}=0.5$, and $J_{z}=2$.}
\end{figure*}

In this section, we will provide two examples to demonstrate the mechanism obtained in Sec.~\ref{Sec2}.
As discussed following, despite these two examples hold far different band structure and symmetric operator, the occupation probability and complete destructive condition all can be predicted by Eq.~\eqref{P-+}.

%%%%%%%%%%%%%%%%%%%%%%%%%%%%%%%%%%%%%%%%%%%%%%%%%%%%%%%%%%%%%%%%%%%%%
\subsection{Example I: system with two sharp band minima}\label{Sec31}
%%%%%%%%%%%%%%%%%%%%%%%%%%%%%%%%%%%%%%%%%%%%%%%%%%%%%%%%%%%%%%%%%%%%%
To begin, we illustrate our mechanism via a concrete example with the symmetric operator $\hat{\sigma}_{r}=\hat{\sigma}_{z}$ ($\theta=\varphi=0$ as defined in Eq.~\eqref{Symmetry}) and take 
\begin{eqnarray}\label{Ham11}
d_{0}(t)&=&J_{0}\cos(\pi t/T),~~~
d_{x}(t)=J_{x}\sin^{2}(\pi t/T),\cr\cr
d_{y}(t)&=&J_{y}\sin^{2}(\pi t/T),~~
d_{z}(t)=J_{z}\cos^{21}(\pi t/T).
\end{eqnarray}
After setting parameters as $J_{0}=1$, $J_{x}=1.5$, $J_{y}=0.5$, and $J_{z}=2$, two sharp band minima appear during the whole evolution as shown in Fig.~\ref{Example1}(a).
As discussed in Sec.~\ref{Sec2} and displayed in Fig.~\ref{Example1}(a), we can divide the whole LZSM interference into three stages: non-adiabatic transition around the first band minimum in stage I, adiabatic dynamical phase accumulation in stage II, another non-adiabatic transition around the second band minimum in stage III.

We start from the ground state $|E_{-}(0)\rangle$ of the TLS at initial time $t=0$.
The occupation probability of state remaining in ground state manifold, $P_{-}(t)=|\langle E_{-}(t)|\Psi(t)\rangle|^2$, approaches to $1$ until evolution arrives near the first band minimum as shown in Fig.~\ref{Example1}(b).
Then the TLS undergoes a non-adiabatic transition and the occupation probability reduces to $P_{-}\approx0.5$ after passing the band minimum region.
Next, the TLS evolves adiabatically In stage II and $P_{-}$ does not vary with evolution time $t$.
Finally, another second non-adiabatic transition occurs in stage III giving rise to final occupation probabilities $P_{T}$ which is heavily dependent on total evolution time $T$.
For instance, the TLS will stay in the ground state for $T=2$ while is almost transition to excited state for $T=8$, as displayed in Fig.~\ref{Example1}(b).
More specifically, we calculate the time derivative $\dot{P}_{-}$ of occupation probabilities $P_{-}$ as shown in Fig.~\ref{Example1}(c), and the end point $t_f$ of stage I is obtained as the first minimum of $\dot{P}_{-}$.
And then the stage II will be the time interval $(t_f,T-t_f)$.

After confirming the end point $t_f$, we can extract the absolute value of reflection and transmission index $|\mathcal{R}|$ and $|\mathcal{T}|$, as well as phase factor $\phi_{\mathcal{L}}$.
In present system, we have $|\mathcal{R}|^2=|\mathcal{T}|^2=0.5$ and $4|\mathcal{R}|^2|\mathcal{T}|^2=1$ in diabatic limit ($T\rightarrow0$) as shown in Fig.~\ref{Example1}(d), because chiral-mirror-like symmetrty in Eq.~\eqref{Symmetry} as well as the relation between initial Hamiltonian and symmetric operator $[\hat{H}(0),\hat{\sigma}_{r}]=0$. 
With the increasing of total evolution time $T$, the opposite trend of $|\mathcal{R}|^2$ (increase, blue solid line) and $|\mathcal{T}|^2$ (decrease, red dashed line) is observed and then $4|\mathcal{R}|^2|\mathcal{T}|^2$ decreases from $1$ slowly.
Meanwhile, the phase factor $\phi_{\mathcal{L}}$ varies with total evolution time $T$ greatly, mainly because of $\phi_{d}$, the dynamical phases accumulating in stage II, as shown in Fig.~\ref{Example1}(e). 
After extracting $4|\mathcal{R}|^2|\mathcal{T}|^2$ and $\phi_{\mathcal{L}}$, the probability $\mathcal{P_{-+}}$ and $\mathcal{P_{--}}$ can be obtained through Eq.~\eqref{P-+}
We focus on the probability $\mathcal{P_{--}}=1-4|\mathcal{R}|^2|\mathcal{T}|^2\cos^2\phi_{\mathcal{L}}$ that is associating with the TLS evolves to ground sates of $\hat{H}(T)$ (also the initial excited state) after LZSM interference.
As shown in Fig.~\ref{Example1}(f), $\mathcal{P_{--}}$ oscillates versus total evolution time $T$ demonstrating the feature of interference and the numerical results (red asterisk) agree well with the TM method solutions (blue solid line). 
In detail, the maxima are kept constant as $1$, independent of $|\mathcal{R}|$ and $|\mathcal{T}|$, only requires phase factor to satisfy $\phi_{\mathcal{L}}=(n+1/2)\pi$ with $n\in\mathbb{N}$ [Figs.~\ref{Example1}(e)-(f)].
Furthermore, we have $\mathcal{P_{--}}=0$ in diabatic limit ($T\rightarrow0$) due to $|E_{+}(T)\rangle=|E_{-}(0)\rangle$ and the value of minima, occurring at $\phi_{\mathcal{L}}=n\pi$ with $n\in\mathbb{N}$, increases with $T$ due to the reduction of $4|\mathcal{R}|^2|\mathcal{T}|^2$ [Figs.~\ref{Example1}(d)-(f)].

We note that, at initial time $t=0$ the ground state $|E_{-}(0)\rangle=|0\rangle$ and the excited state $|E_{+}(0)\rangle=|1\rangle$ are the eigenstates of $\sigma_z$.
Considering the commutate relationship $[\hat{H}(0),\hat{\sigma}_{r}]=0$ discussed above, we can achieve quantum control of single-qubit state by adjusting the total evolution time $T$ as well as $\phi_{\mathcal{L}}$.
In the case $\phi_{\mathcal{L}}=(n+1/2)\pi$ with $n\in\mathbb{N}$, the system undergoes non-adiabatic rapid passage and ends up in state $|1\rangle$.

%%%%%%%%%%%%%%%%%%%%%%%%%%%%%%%%%%%%%%%%%%%%%%%%%%%%%%%%%%%%%%%%%%%%%
\subsection{Example II: system with flat bands and holding time}\label{Sec32}
%%%%%%%%%%%%%%%%%%%%%%%%%%%%%%%%%%%%%%%%%%%%%%%%%%%%%%%%%%%%%%%%%%%%%
\begin{figure}[!htp]
\includegraphics[width=1\columnwidth]{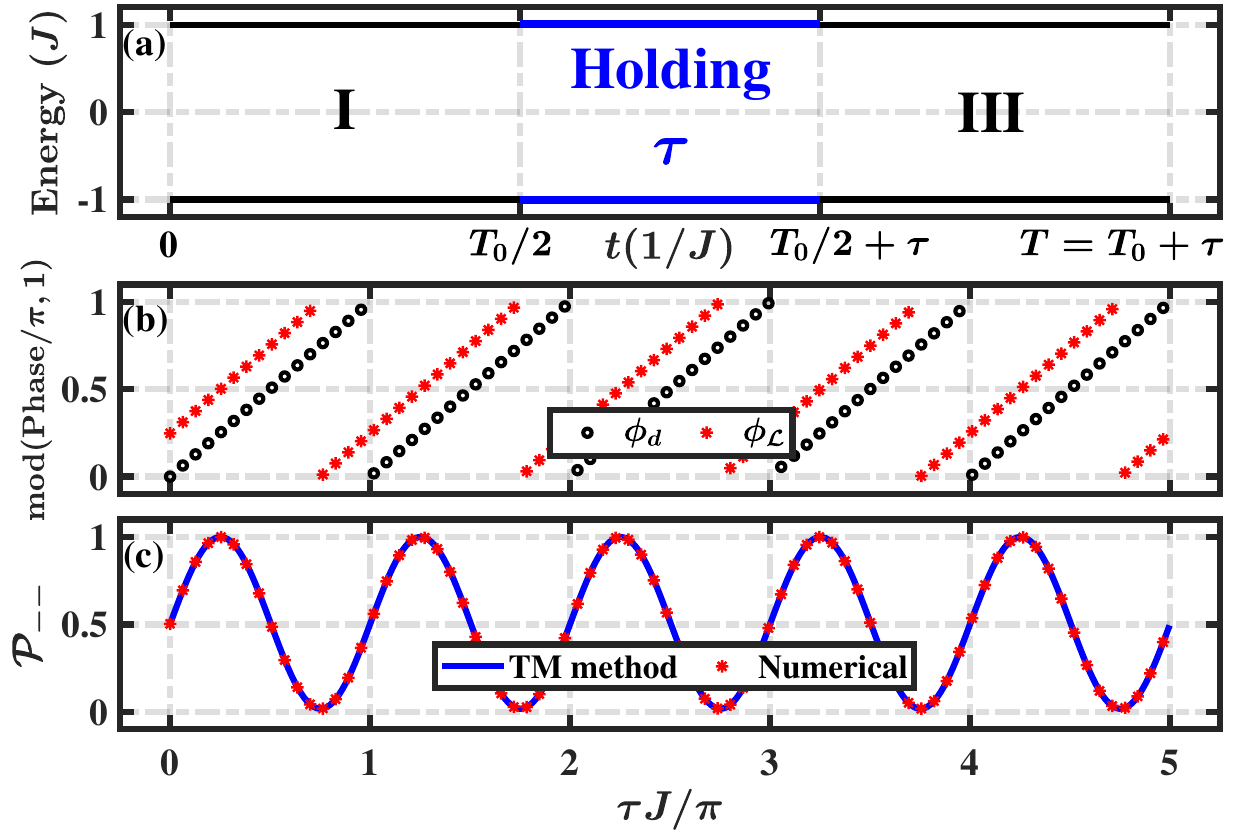}
\caption{\label{Example2}LZSM interference of TLS with flat bands and holding time. (a) Energy spectrum of Hamiltonian defined by Eqs.~\eqref{Ham2} and \eqref{Ham12}. The energy bands are flat during the whole evolution. The end point is $t_f=T_0/2$ for stage I and the total evolution time is $T=T_0+\tau$. (b) and (c) LZSM interference results starting from ground state $|E_{-}\rangle$ versus different holding duration $\tau$ and constant normal evolution duration $T_0=1.26$. (b) Phase factor $\phi_{d}=\tau$ (black circle) and $\phi_{\mathcal{L}}$ (red asterisk). The phase factor from parallel transport gauge is $2\phi_{c}=0.282\pi$ while the relative phase between the reflection coefficient $\mathcal{R}$ and the transmission coefficient $\mathcal{T}$ at $t_f$ is $\phi_{r}=0.966\pi$. (c) Transfer matrix (TM) method solutions (blue solid line) obtained in Eq.~\eqref{P-+} and numerical results (red asterisk) of occupation probability $\mathcal{P_{--}}$. It oscillates periodically over holding duration $\tau$ with a period of $\tau J=\pi$. In all panels, parameters are set as $\theta=\pi/3$, $\varphi=\pi/6$, $J_{0}=1$, $J_{x}=1.5$, $J_{y}=0.5$, and $J_{z}=2$.}
\end{figure}
To proceed further, we consider a TLS with flat bands and accumulate dynamical phase $\phi_{d}$ by holding the system for a variable duration $\tau$ in this subsection. 
Without loss of generality, we chose the symmetric operator $\hat{\sigma}_{r}=\frac{3}{4}\hat{\sigma}_{x}+\frac{\sqrt{3}}{4}\hat{\sigma}_{y}+\frac{1}{2}\hat{\sigma}_{z}$ ($\theta=\pi/3$ and $\varphi=\pi/6$ as defined in Eq.~\eqref{Symmetry}) and take
\begin{equation}
d_\theta=\left\{
\begin{aligned}
&J_\theta\sin(\pi t/T_0)&0\leq t \leq T_0/2\\
&J_\theta&T_0/2< t < T_0/2+\tau\\
&J_\theta\sin\big[\pi (t-\tau)/T_0\big]&T_0/2+\tau \leq t \leq T
\end{aligned}
\right.\nonumber
\end{equation}
\begin{equation}\label{Ham12}
d_r=\left\{
\begin{aligned}
&J_r\cos(\pi t/T_0)&0\leq t \leq T_0/2\\
&0&T_0/2< t < T_0/2+\tau\\
&J_r\cos\big[\pi (t-\tau)/T_0\big]&T_0/2+\tau \leq t \leq T
\end{aligned}
\right.
\end{equation}
\begin{equation}
d_\varphi=\left\{
\begin{aligned}
&J_\varphi\sin(\pi t/T_0)&0\leq t \leq T_0/2\\
&J_\varphi&T_0/2< t < T_0/2+\tau\\
&J_\varphi\sin\big[\pi (t-\tau)/T_0\big]&T_0/2+\tau \leq t \leq T
\end{aligned}
\right.\nonumber
\end{equation}

We plot the energy spectrum versus evolution time $t$ in Fig.~\ref{Example2}(a) where the parameters are set as $J_{\theta}=\sqrt{2}/2$, $J_{r}=1$, and $J_{\varphi}=-\sqrt{2}/2$,
As demonstrated here, two separated energy bands are completely flat ($E_{\pm}/J=\pm1$) during the whole evolution.
The TLS undergoes two LZ transitions in stages I and III.
These transitions are constrained by chiral-mirror-like symmetry and are separated by an intermediate holding duration $\tau$, which controls the accumulation of the dynamical phase $\phi_{d}$.
Here, the end point for stage I is represented by $t_f=T_0/2$ and then the total evolution time will be $T=T_0+\tau$, as shown in Fig.~\ref{Example2}(a).
Specifically, we study the LZSM interference with differen holding duration $\tau$ and unchanged normal evolution duration $T_0=1.26$.
With these settings, the phase factors $\phi_{c}$ and $\phi_{r}$ as well as $4|\mathcal{R}|^2|\mathcal{T}|^2$ are irrelevant to $\tau$, which are obtained as $2\phi_{c}=0.282\pi$ and $\phi_{r}=0.966\pi$ and $4|\mathcal{R}|^2|\mathcal{T}|^2\approx1$.
Meanwhile, the dynamical phase $\phi_{d}$ accumulating during the holding duration $\tau$ will take simple form, i.e., $\phi_{d}=\tau$ resulting to phase factor $\phi_{\mathcal{L}}$ increasing linearly with holding duration as $2\phi_{c}+\phi_{r}+\tau$, as shown in Fig.~\ref{Example2}(b).
Therefore, the destructive and constructive patterns of LZSM interference can be predicted through the occupation probability $\mathcal{P_{--}}$ versus $\tau$.
As demonstrated in fig.~\ref{Example1}(c), probability $\mathcal{P_{--}}$ oscillates between $0$ and $1$ periodically over holding duration $\tau$ with an period of $\tau J=\pi$ and the numerical results (red asterisk) agree well with the TM solutions (blue solid line) obtained in Eq.~\eqref{P-+}. 
%

%%%%%%%%%%%%%%%%%%%%%%%%%%%%%%%%%%%%%%%%%%%%%%%%%%%%%%%%%%%%%%%%%%%%%%%%%%%%%%%%%%%%
\section{Non-adabatic topological transport of edge states}\label{Sec4}
%%%%%%%%%%%%%%%%%%%%%%%%%%%%%%%%%%%%%%%%%%%%%%%%%%%%%%%%%%%%%%%%%%%%%%%%%%%%%%%%%%%%
\begin{figure}[!htp]
\includegraphics[width=1\columnwidth]{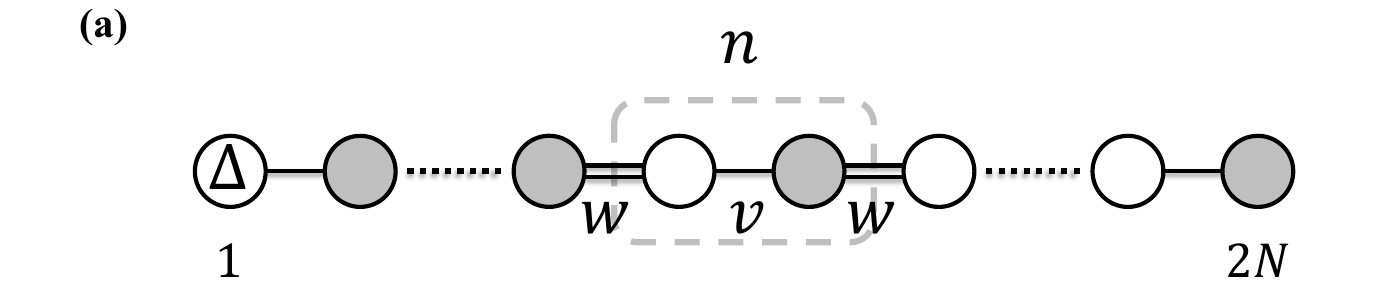}
\includegraphics[width=1\columnwidth]{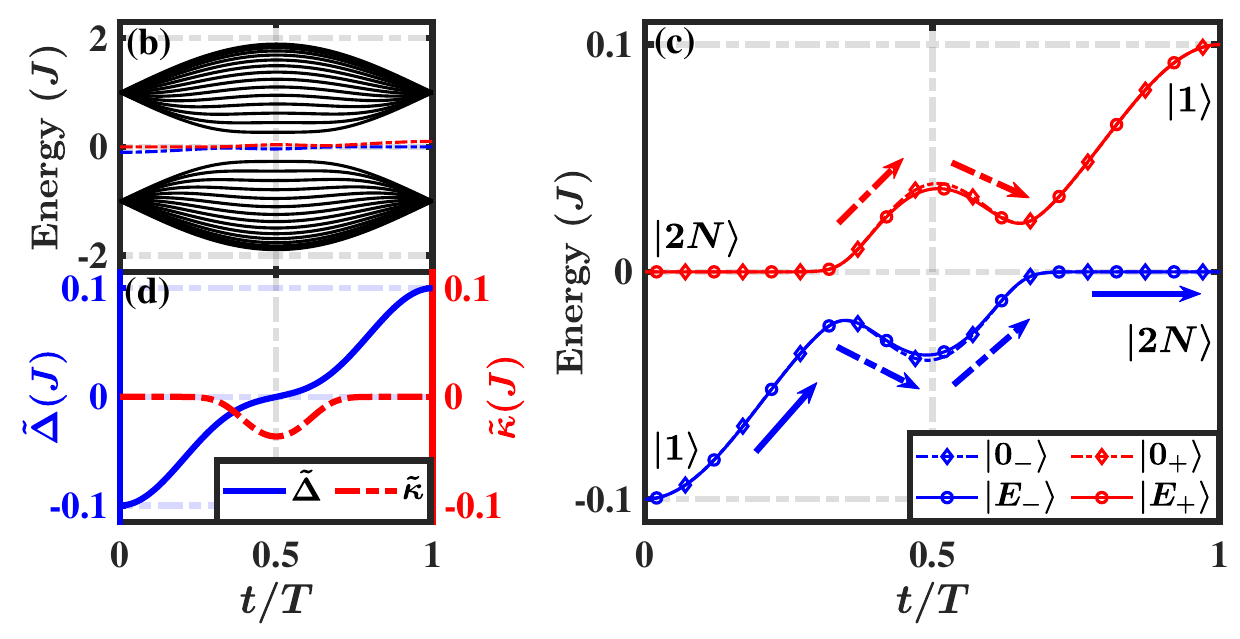}
\includegraphics[width=1\columnwidth]{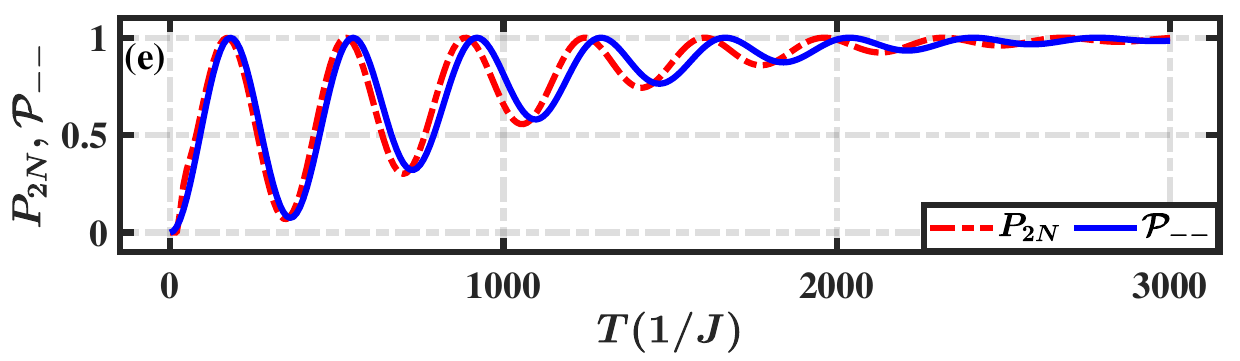}
\caption{\label{SSHModel}Non-adiabatic topological transport of edge states. (a) Schematic diagram of the SSH model. Empty (filled) circles are odd (even) sites, which are grouped into unit cells. The $n$-th cell is circled by a dotted line. $\Delta$ is the local on-site energy of site $1$, $v$ and $w$ denote the intracell and intercell hopping, respectively.
(b) Energy spectrum for a finite-size SSH model described by Hamiltonian in eq.~\eqref{HamSSH}. Black solid lines represent the bulk states while blue and red dashed lines represent the edge states $|0_{-}\rangle$ and $|0_{+}\rangle$, respectively.
(c) Energy spectrum for edge states, $|0_{-}\rangle$ (blue dashed line with diamonds) and $|0_{+}\rangle$ (red dashed line with diamonds), and reduced two-level model described in Eq.~\eqref{HamR} ground state $|E_{-}\rangle$ (blue solid line with circles) and excited state $|E_{+}\rangle$ (red solid line with circles).
(d) Effective coefficients $\tilde{\Delta}$ (blue solid line) and $\tilde{\kappa}$ (red dashed line) in Eq.~\eqref{coefficients} of reduced two-level model. Here $\tilde{\Delta}(t)=-\tilde{\Delta}(T-t)$ and $\tilde{\kappa}(t)=\tilde{\kappa}(T-t)$ ensure the chiral-mirror-like symmetry of the reduced two-level model.
(e) Occupation probability $P_{2N}$ (red dashed line) starts from $|0_{-}\rangle$ for the SSH model and $\mathcal{P_{--}}$ (blue solid line) starts from $|E_{-}\rangle$ versus total evolution time $T$. In all panels, system contains $2N=32$ sites while parameters are $w=1$, $v_{0}=0.9$ and $\Delta_{0}=0.1$.}
\end{figure}
In the last decade, numerous adiabatic topological transport protocols have been proposed to transport particle or quantum state by adiabatically varying the Hamiltonian of topological states in time.
However, the transport speed is always limited by the adiabatic condition in these works.
In this section, we explore how to achieve non-adiabatic topological transport of edge states, edge-to-edge transport via topological edge channels beyond the adiabatic condition, based on our mechanism of chiral–mirror-like symmetry-protected LZSM interference.
We consider the SSH model~\cite{WPSuPRL1979}, one of the most prototypical models exhibiting topological edge states, as a concrete example, although this mechanism can easily be implemented in other topological models. 
However, the transition between the two eigenenergy branches, mediated by hybridized edge states, is forbidden due to inversion symmetry, as discussed in Ref.~\cite{SHuPRA2024}.
Hence, in addition to tuning the strength of the intracell hopping, we also introduce a time-dependent local on-site energy at the left end to break inversion symmetry. 
The resulting total Hamiltonian is ($\hbar=1$)
\begin{eqnarray}\label{HamSSH} 
\mathcal{\hat{H}}(t)/J&=&\hat{H}_{\rm{SSH}}(t)+\hat{H}_{\Delta}(t),\cr\cr
\hat{H}_{\rm{SSH}}(t)&=&v(t)\sum_{n=1}^{N}\hat{b}_{2n-1}^{\dag}\hat{b}_{2n}
+w\sum_{n=1}^{N-1}\hat{b}_{2n}^{\dag}\hat{b}_{2n+1}+{\rm H.c.},\cr\cr\cr
\hat{H}_{\Delta}(t)&=&\Delta(t)\hat{b}_{1}^{\dag}\hat{b}_{1}.
\end{eqnarray}
Here $N$ is total unit cells (the total number of lattice sites is $2N$), $\hat{b}_n^{\dag}$ and $\hat{b}_n$ are creation and annihilation operators acting on lattice site $n$, respectively, $v(t)=v_{0}\sin(\pi t/T)$ and $w=1$ represent the intracell and intercell hopping amplitudes, and $\Delta(t)=-\Delta_{0}\cos(\pi t/T)$ represents the on-site energy at the left end of the SSH model, as shown in Fig.~\ref{SSHModel}(a).

We plot the energy spectrum versus evolution time $t$ for a SSH chain with $2N=32$ sites in Fig.~\ref{SSHModel}(b) with parameters being $w=1$, $v_{0}=0.9$, and $\Delta_{0}=0.1$.
During the whole evolution process, the system always stays in the topological region and the in-gap topological edge states $|0_{-}\rangle$ (lower, blue dashed line) and $|0_{+}\rangle$ (up, red dashed line) are well separated from the bulk states (black solid line). 
As a result, we can only focus on the edge states manifold during the evolution.
Obviously, the adiabatic topological transport will cost a lot of evolution time because the energy gap between the edge states is really small.
However, the non-adiabatic topological transport with a short evolution time can be constructed as explained below.

To be specific, the SSH chain is in the fully dimerized limit at initial time $t=0$ and the edge states are $|0_{-}\rangle=|1\rangle$ and $|0_{+}\rangle=|2N\rangle$ as shown in Fig.~\ref{SSHModel}(c).
Therefore, a particle initially injected in site $1$ ($|1\rangle$) will occupy the edge state $|0_{-}\rangle$.
Next, the path of the particle is split around the first minimum of the energy gap due to non-adiabatic effect and then merges at the second one as illustrated in Fig.~\ref{SSHModel}(c). 
As a result, the final occupation probability of the particle at $t=T$ depends not only on the transition process around the two symmetric energy gap minima but also the phase accumulation process between them. 
Hence, this physical scenario is able to realize the LZSM interference of topological edge states. 

Moreover, the SSH chain returns to the fully dimerized limit at the final time $t=T$.
However, the edge states exchange their forms, i.e. $|0_{-}(0)\rangle=|0_{+}(T)\rangle=|1\rangle$ and $|0_{+}(0)\rangle=|0_{-}(T)\rangle=|2N\rangle$, due to the local on-site energy taking odd function form $\Delta(t)=-\Delta(T-t)$.
Therefore we focus on the probability that the particle remains in the lower branch of edge state $|0_{-}\rangle$ after LZSM interference, which corresponds to topological transport of the particle from one end to the other quantified by $P_{2N}=|\langle 2N|\Psi(T)\rangle|^2$. 
Fig.~\ref{SSHModel}(e) plots $P_{2N}=|\langle 2N|\Psi(T)\rangle|^2$ versus total evolution time $T$ as blue solid line, where the oscillating behavior is observed.
Remarkably, while the value of minimum increases with $T$ and reaches $1$ in the adiabatic limit $T\rightarrow\infty$, these maxima are kept constant as $1$ same to the symmetry-protected LZSM interference for TLS discussed in Sec.~\ref{Sec31}.
To sum up, the non-adiabatic topological transport is realized with complete destructive LZSM interference and is far more efficient with the first maximum of $P_{2N}=1$ at $T=170$ while it requires $T>3000$ to satisfy adiabatic condition.

In order to gain a better understanding of the mechanism inducing topological transport, we first provide the form of topological edge states as follows:
\begin{eqnarray}\label{EdgeState} 
|L\rangle&=&|1\rangle+\eta|3\rangle+\cdots+\eta^{N-1}|2N-1\rangle,\cr\cr
|R\rangle&=&|2N\rangle+\eta|2N-2\rangle+\cdots+\eta^{N-1}|2\rangle,
\end{eqnarray}
which are obtained in thermodynamic limit with $\eta=-v(t)/w$ being the localization factor and only localize in one site under dimerized limit ($t=0,T$), i.e., $|L\rangle=|1\rangle$ and $|R\rangle=|2N\rangle$.
Next, we derive an effective TLS described by reduced Hamiltonian $\mathcal{\hat{H}_{R}}$ under base vectors $|L\rangle$ and $|R\rangle$:
\begin{eqnarray}\label{HamR} 
\mathcal{\hat{H}_{R}}(t)/J
&&=\left(
\begin{array}{llllllllll}
\tilde{\Delta}(t)~~\tilde{\kappa}(t)\cr
~\tilde{\kappa}(t)~~~~0
\end{array}\right)\cr\cr\cr
&&=\frac{\tilde{\Delta}(t)}{2}\hat{\sigma}_{0}
+\tilde{\kappa}(t)\hat{\sigma}_{x}
+\frac{\tilde{\Delta}(t)}{2}\hat{\sigma}_{z},
\end{eqnarray}
with
\begin{eqnarray}\label{coefficients}
\tilde{\Delta}(t)
&\equiv&\langle{L}|\mathcal{\hat{H}}|L\rangle
=\frac{\eta^{2}-1}{\eta^{2N}-1}\Delta(t),\cr\cr
\tilde{\kappa}(t)
&\equiv&\langle{L}|\mathcal{\hat{H}}|R\rangle
=\langle{R}|\mathcal{\hat{H}}|L\rangle
=\frac{\eta^{N-1}(\eta^{2}-1)}{\eta^{2N}-1}v(t),
\end{eqnarray}
as plotted in Fig.~\ref{SSHModel}(d).
Interestingly, these effective coefficients satisfy $\tilde{\Delta}(t)=-\tilde{\Delta}(T-t)$ and $\tilde{\kappa}(t)=\tilde{\kappa}(T-t)$, ensuring that this TLS preserves chiral-mirror-like symmetry with $\mathcal{\hat{H}_{R}}(t)=-\hat{\sigma}_{z}\mathcal{\hat{H}_{R}}(T-t)\hat{\sigma}_{z}$ same to the one discussed in Sec.~\ref{Sec31}.
Meanwhile, the commutation relation, $[\mathcal{\hat{H}_{R}}(0),\hat{\sigma}_{z}]=0$, ensures the fact $|E_{-}(T)\rangle=|E_{+}(0)\rangle=|R\rangle$ and $|E_{+}(T)\rangle=|E_{-}(0)\rangle=|L\rangle$.
Moreover, its energy spectrum (blue and red solid lines with circles) also exhibits two sharp band minima as shown in Fig.~\ref{SSHModel}(c).
Then the symmetry-protected LZSM interference will occur where the complete destructive result $\mathcal{P_{--}}=1$ associates with the process of topological transport from left edge state $|L\rangle$ to right one $|R\rangle$.
In Fig.~\ref{SSHModel}(e), we plot $\mathcal{P_{--}}$ versus total evolution time $T$ (blue solid line) and its feature are as same as $P_{2N}$.
For current result with $2N=32$, the time $T=181.4$ is associated with the first peak that realizes perfect transport (effective TLS).
As a comparison, the quantum speed limit (QSL), which is due to energy-time uncertainty~\cite{Mandelstam1945,Margolus1998}, will give a time $T_{\rm{QSL}}=180$.
Here the QSL is obtained via $T_{\rm{QSL}}=\frac{\pi}{2\langle\tilde{\kappa}(t)\rangle}$, where $\langle\tilde{\kappa}(t)\rangle$ is the time average value of coupling strength~\cite{CanevaPRL2009,MalossiPRA2013}.
Thereby, our protocol could be applied to fast topological quantum state transfer close to QSL via spin chain.
However, because the tiny energy difference between the SSH model and the reduced TLS especially the energy gap of the latter is smaller in adiabatic region, dynamical phase $\phi_d$ obtained for TLS is smaller than the one for SSH model with same evolution time.
As a result, $P_{2N}$ takes short total evolution time $T$ to reach its maximum value comparing to $\mathcal{P_{--}}$ and their difference may grow with $T$, as predicted by Eq.~\eqref{P-+} and shown in Fig.~\ref{SSHModel}(e). 
Combining these results, we have constructed the non-adiabatic topological transport and confirmed it is rooted in the symmetry-protected LZSM interference of topological edge states.
This approach can be applied to speed up the braiding of Majorana fermions, which involves the topological transport and was always handled adiabatically~\cite{AliceaNP2011,ElliottRMP2015}.

%%%%%%%%%%%%%%%%%%%%%%%%%%%%%%%%%%%%%%%%%%%%%%%%%%%%%%%%%%%%%%%%%%%%%%%%%%%%%%%%%%%%%%%%%%%%%%%%%%%%%%%%%%%%%%%%%%%%%%%%%%%%%%%%%%%%%%%
\section{SUMMARY AND DISCUSSION}\label{Sec5}
%%%%%%%%%%%%%%%%%%%%%%%%%%%%%%%%%%%%%%%%%%%%%%%%%%%%%%%%%%%%%%%%%%%%%%%%%%%%%%%%%%%%
In this paper, we systematically investigated the LZSM interference under chiral-mirror-like symmetry and calculated the occupation probability based on the TM method.
It has been shown that complete destructive interference can be observed in chiral-mirror-like symmetric systems, which cover a wide range of systems benefiting from the relative arbitrariness of the symmetric operator.
To illustrate these results, two TLSs are presented as concrete examples.
Although they exhibit different characteristics and are constrained by distinct symmetric operators, the symmetry-protected LZSM interference accompanied by rapid state transitions can be observed.
As a nontrivial application of this mechanism, we have proposed a non-adiabatic topological transport of edge states in the SSH chain, going beyond conventional adiabatic approaches.
Moreover, we point out that our scheme can be extended to non-Hermitian systems~\cite{WangNJP2023,ShenPRA2019,BergholtzRMP2021,LeiCP2024}.

Lastly, we briefly discuss the experimental feasibility of our mechanism. 
The TLS is one of the basic models in quantum physics, taking into account that various physical systems can be described by a two-level model. 
LZ transition and LZSM interference have been observed in a wide range of physical systems, including ultracold molecules~\cite{MarkPRL2007}, quantum dots~\cite{CaoNC2013}, nanostructures~\cite{AverinPRL1995,AverinPRL1999}, Bose-Einstein condensates~\cite{ZenesiniPRL2009}, Rydberg atoms~\cite{ZhangPRL2018}, atomic qubits~\cite{FerrierPRL2013}, graphene~\cite{HiguchiNature2017}, and waveguide systems~\cite{ChenPRL2021,XiePRB2022} etc.
In principle, the mechanism is robust against symmetry preserved noise.
Even more, when part of noises deviate the form required by symmetry, our mechanism still works if their average effects vanish~\cite{SHuPRA2024}.
Thus, testing our mechanism in such systems is convenient.
On the other hand, photonic waveguide arrays represent a promising platform for exploring topological features.
Adiabatic control of topological edge states, including non-quantized adiabatic pumping of topological edge state~\cite{YEKrausPRL2012} and quantum interference of topological states of light~\cite{JLTambascoSA2018},  have been observed on this system.
Modulated photonic waveguide arrays may serve as a potential platform for implementing non-adiabatic topological transport based on symmetry-protected LZSM interference.

\acknowledgements{This work is supported by the National Natural Science Foundation of China under Grant No. 12104103, the Guangdong Basic and Applied Basic Research Foundation under Grant No. 2022A1515010726, Science and Technology Program of Guangzhou under Grant No. 2023A04J0039, and Research capacity improvement project of doctoral program construction unit of Guangdong Polytechnic Normal University Grant No. 22GPNUZDJS32.}

\appendix
\renewcommand\theequation{\Alph{section}\arabic{equation}}

%%%%%%%%%%%%%%%%%%%%%%%%%%%%%%%%%%%%%%%%%%%%%%%%%%%%%%%%%%%%%%%%%%%%%%%%%%%%%%%%%%%%
\section{The symmetric relation between two non-adiabatic transition matrices}\label{appA}
%%%%%%%%%%%%%%%%%%%%%%%%%%%%%%%%%%%%%%%%%%%%%%%%%%%%%%%%%%%%%%%%%%%%%%%%%%%%%%%%%%%%
In this section, we will discuss the relation between $\hat{U}_{\rm{I}}$ and $\hat{U}_{\rm{III}}$, and the choice of gauge in calculating them. 
Obviously, under diabatic basis $\{|0\rangle,|1\rangle\}$, the chiral-mirror-like symmetry will enforce non-adiabatic transition matrices in stages I and III holding a relation: 
\begin{eqnarray}\label{U13APP}
\tilde{U}_{\rm{I}}=\hat{\sigma}_{r}\tilde{U}^{-1}_{\rm{III}}\hat{\sigma}_{r}.
\end{eqnarray}
However we need to express this relation in $\{|E_{-}(t)\rangle,|E_{+}(t)\rangle\}$ basis, and then the investigation of parallel transport gauge is necessary.

Firstly, we define a set of eigenstates $|\mathcal{E}_{\pm}(t)\rangle$ with smooth gauge and require them satisfying
\begin{eqnarray}
|\mathcal{E}_{\pm}(t)\rangle
=\hat{\sigma}_{r}|\mathcal{E}_{\mp}(T-t)\rangle.
\end{eqnarray} 
In the time point $T/2$, we have 
\begin{eqnarray}\label{UThalfApp}
[\hat{H}(T/2),\hat{\sigma}_{r}]_+=0,~~ |\mathcal{E}_{+}(T/2)\rangle=\hat{\sigma}_{r}|\mathcal{E}_{-}(T/2)\rangle. 
\end{eqnarray}
Based on the transformation: 
\begin{eqnarray}\label{SrTranApp}
\hat{U}_{r}\hat{\sigma}_{r}\hat{U}^{-1}_{r}&=&\hat{\sigma}_z,~~~~ 
\hat{U}_{r}\hat{\sigma}_{\theta}\hat{U}^{-1}_{r}=\hat{\sigma}_x,\cr\cr
\hat{U}_{r}\hat{\sigma}_{\varphi}\hat{U}^{-1}_{r}&=&\hat{\sigma}_y,~~~~
\hat{U}_{r}=e^{i\theta\hat{\sigma}_{y}/2}
e^{i\varphi\hat{\sigma}_{z}/2},
\end{eqnarray}
we have
\begin{eqnarray}
[\hat{U}_{r}\hat{H}(T/2)\hat{U}^{-1}_{r},\hat{\sigma}_z]_+=0,
\end{eqnarray}
and then $\hat{U}_{r}|\mathcal{E}_{\pm}(T/2)\rangle$ will be eigenstate of $\hat{U}_{r}\hat{H}(T/2)\hat{U}^{-1}_{r}$ satisfying 
\begin{eqnarray}
\hat{U}_{r}|\mathcal{E}_{+}(T/2)\rangle=
\hat{\sigma}_z\hat{U}_{r}|\mathcal{E}_{-}(T/2)\rangle.
\end{eqnarray}
Thereby the ground eigenstates of $\hat{U}_{r}\hat{H}(T/2)\hat{U}^{-1}_{r}$ will be
\begin{eqnarray}
\hat{U}_{r}|\mathcal{E}_{-}(T/2)\rangle
=\frac{1}{\sqrt{2}}\left(
\begin{array}{llllllllll}
e^{-i\vartheta/2}\cr
-e^{i\vartheta/2}
\end{array}\right),
\end{eqnarray}
with $\vartheta\equiv\arg[d_{\theta}(T/2)-id_{\varphi}(T/2)]$.
And another eigenstate will be
\begin{eqnarray}
\hat{U}_{r}|\mathcal{E}_{+}(T/2)\rangle=\hat{\sigma}_z\hat{U}_{r}|\mathcal{E}_{-}(T/2)\rangle
=\frac{1}{\sqrt{2}}\left(
\begin{array}{llllllllll}
e^{-i\vartheta/2}\cr
e^{i\vartheta/2}
\end{array}\right).\nonumber\\
\end{eqnarray}
Then the eigenstates of $\hat{H}(T/2)$ will be 
\begin{widetext}
\begin{eqnarray*}\label{EThalfApp}
|\mathcal{E}_{-}(T/2)\rangle=
\big[\cos(\theta/2)e^{-i(\vartheta+\varphi)/2}
+\sin(\theta/2)e^{i(\vartheta-\varphi)/2},~~
\sin(\theta/2)e^{-i(\vartheta-\varphi)/2}
-\cos(\theta/2)e^{i(\vartheta+\varphi)/2}\big]^T/\sqrt{2},\nonumber\\
|\mathcal{E}_{+}(T/2)\rangle=
\big[\cos(\theta/2)e^{-i(\vartheta+\varphi)/2}
-\sin(\theta/2)e^{i(\vartheta-\varphi)/2},~~
\sin(\theta/2)e^{-i(\vartheta-\varphi)/2}
+\cos(\theta/2)e^{i(\vartheta+\varphi)/2}\big]^T/\sqrt{2}. \end{eqnarray*}
\end{widetext}
From the above equation, we can get 
\begin{eqnarray}\label{ERealApp}
\langle\mathcal{E}_{+}(T/2)|0\rangle\langle0|\mathcal{E}_{-}(T/2)\rangle=\cos\theta+i\sin\theta\sin\vartheta,\nonumber\\
\end{eqnarray}
which holds phase factor 
$2\phi_{ c}\equiv\arg(\cos\theta+i\sin\theta\sin\vartheta)$.
Therefore, it is reasonable to require this eigenstate basis at the first half of the process satisfying 
\begin{eqnarray}\label{EconFHApp}
\arg\big[\langle\mathcal{E}_{+}(t)|0\rangle
\langle0|\mathcal{E}_{-}(t)\rangle\big]
=2\phi_{c},~~t\in[0,T/2]. 
\end{eqnarray}
In addition, if we require transferring matrix $\mathcal{S}(t)=\big[|\mathcal{E}_{-}(t)\rangle,
|\mathcal{E}_{+}(t)\rangle\big]$ is unitary and unimodular, these eigenstates with $t\in[0,T/2]$ will be
\begin{eqnarray}\label{EconFHAppC}
|\mathcal{E}_{-}(t)\rangle&=&e^{i\phi_{c}}
\left(
\begin{array}{llllllllll}
\cos\eta(t)e^{-i\chi(t)}\cr
-\sin\eta(t)e^{i\chi(t)}
\end{array}
\right),\cr\cr
|\mathcal{E}_{+}(t)\rangle&=&e^{-i\phi_{c}}
\left(
\begin{array}{llllllllll}
\sin\eta(t)e^{-i\chi(t)}\cr
\cos\eta(t)e^{i\chi(t)}
\end{array}
\right).
\end{eqnarray}
What's more, the eigenstate basis at the second half of the process can be obtained through symmetry relation 
\begin{eqnarray}\label{EconLHApp}
|\mathcal{E}_{\pm}(t)\rangle
=\hat{\sigma}_{r}|\mathcal{E}_{\mp}(T-t)\rangle,~~t\in[T/2,T]. 
\end{eqnarray}  

Next, we define $|E_{\pm}(t)\rangle=e^{i\gamma_{\pm}(t)}|\mathcal{E}_{\pm}(t)\rangle$ with $\gamma_{\pm}(T/2)=0$, which holds parallel transport gauge and used in main text, i.e., 
\begin{eqnarray}\label{TranguaApp}
\langle E_{\pm}(t)|\frac{\partial}{\partial t}|E_{\pm}(t)\rangle=0,
\end{eqnarray}  
giving rise to the limitation of phase factor $\gamma_{\pm}(t)$ during the first half of the process
\begin{eqnarray}\label{TranguaApp}
\partial_t\gamma_{-}(t)=-\partial_t\gamma_{+}(t)=\partial_t\chi(t)\cos[2\eta(t)],~~t\in[0,T/2].\nonumber\\
\end{eqnarray}
And the phase factor $\gamma_{\pm}(t)$ during the second half of the process satisfies 
\begin{eqnarray}\label{EanalinAPP}
\partial_t\gamma_{-}(t)&=&
-\partial_t\gamma_{+}(t)
=\langle \mathcal{E}_{-}(t)|\partial_t|\mathcal{E}_{-}(t)\rangle\cr\cr
&=&\langle \mathcal{E}_{+}(T-t)|\hat{\sigma}_{\mathbf{r}}\partial_t\hat{\sigma}_{\mathbf{r}}|\mathcal{E}_{+}(T-t)\rangle \cr\cr
&=&-\partial_t\gamma_{-}(T-t),
~~~~~~~~t\in[T/2,T].
\end{eqnarray} 
With the boundary condition $\gamma_{\pm}(T/2)=0$, we have 
\begin{eqnarray}\label{FaseAPP}
\gamma(t)\equiv\gamma_{-}(t)=-\gamma_{+}(t)=\gamma_{-}(T-t).
\end{eqnarray} 

We can define unitary and unimodular transformation matrices $\hat{\mathcal{S}}(t)$ and $\hat{S}(t)$:
\begin{eqnarray}\label{TransSAPP}
\hat{\mathcal{S}}(t)=\big[|\mathcal{E}_{-}(t)\rangle,|\mathcal{E}_{+}(t)\rangle\big],~~~ \hat{S}(t)=\big[|{E}_{-}(t)\rangle,|{E}_{+}(t)\rangle\big],\nonumber\\
\end{eqnarray}
transforming non-adiabatic transition matrices under diabatic basis to eigenstates basis:
\begin{eqnarray}\label{Ubas13APP}
\hat{\mathcal{U}}_{\rm{I}}
&=&\hat{\mathcal{S}}^{-1}(t_f)\tilde{U}_{\rm{I}}\hat{\mathcal{S}}(t_i),~~
\hat{\mathcal{U}}_{\rm{III}}
=\hat{\mathcal{S}}^{-1}(T-t_i)\tilde{U}_{\rm{III}}\hat{\mathcal{S}}(T-t_f),\cr\cr
\hat{U}_{\rm{I}}
&=&\hat{S}^{-1}(t_f)\tilde{U}_{\rm{I}}{S}(t_i),~~
\hat{U}_{\rm{III}}=\hat{S}^{-1}(T-t_i)\tilde{U}_{\rm{III}}\hat{S}(T-t_f).\nonumber\\
\end{eqnarray}
From Eq.~\eqref{EconLHApp}, we can know transformation matrices $\hat{\mathcal{S}}(t)$ at two halves of process must satisfy:
\begin{eqnarray}\label{Ubas13ReApp}
\hat{\mathcal{S}}(t)=\hat{\sigma}_{r}\hat{\mathcal{S}}(T-t)\hat{\sigma}_x.
\end{eqnarray}
Furthermore, according to above discussion about phase factor $\gamma$ in Eq.~\eqref{FaseAPP}, these two set of transformation matrices will satisfy:
\begin{eqnarray}\label{Ubas13ReApp}
\hat{S}(t)=\hat{\mathcal{S}}(t)e^{i\gamma(t)\hat{\sigma}_{z}},
\end{eqnarray}
Thereby we have: 
\begin{eqnarray}\label{Ubas1RAPP}
\hat{\mathcal{U}}_{\rm{III}}
&=&\hat{\mathcal{S}}^{-1}(T-t_i)\tilde{U}_{\rm{III}}
\hat{\mathcal{S}}(T-t_f)\cr\cr
&=&\hat{\sigma}_x\hat{\mathcal{S}}^{-1}(t_i)
\hat{\sigma}_{r}\hat{\sigma}_{r}\tilde{U}^{-1}_{\rm{I}}
\hat{\sigma}_{r}\hat{\sigma}_{r}\hat{\mathcal{S}}(t_f)
\hat{\sigma}_x\cr\cr
&=&\hat{\sigma}_x\hat{\mathcal{U}}^{-1}_{\rm{I}}\hat{\sigma}_x
\end{eqnarray}
So after supposing
\begin{eqnarray}
\hat{\mathcal{U}}_{\rm{I}}=\left(
\begin{array}{llllllllll}
\mathcal{R}~~~~-\mathcal{T}^*  \\
\mathcal{T}~~~~~\mathcal{R}^*
\end{array}\right), 
\end{eqnarray}
with $|\mathcal{R}|^2+|\mathcal{T}|^2=1$, we have 
\begin{eqnarray}\label{Ubas3RAPP}
\hat{\mathcal{U}}_{\rm{III}}=
\left(
\begin{array}{llllllllll}
\mathcal{R}~~~~-\mathcal{T}\\
\mathcal{T}^*~~~~~\mathcal{R}^*
\end{array}
\right).
\end{eqnarray}
So the total effect of this LZSM interference will be 
\begin{eqnarray}\label{Ubas13R2App}
\hat{U}&=&\hat{U}_{\rm{III}}\hat{U}_{\rm{II}}\hat{U}_{\rm{I}}\cr\cr
&=&e^{-i\gamma(t_i)\hat{\sigma}_{z}}\hat{\mathcal{U}}_{\rm{III}}
e^{i\gamma(t_f)\hat{\sigma}_{z}}e^{i\phi_{d}\hat{\sigma}_{z}}
e^{-i\gamma(t_f)\hat{\sigma}_{z}}\hat{\mathcal{U}}_{\rm{I}}
e^{i\gamma(t_i)\hat{\sigma}_{z}}\cr\cr
&=&e^{-i\gamma(t_i)\hat{\sigma}_{z}}
\hat{\mathcal{U}}e^{i\gamma(t_i)\hat{\sigma}_{z}},
\end{eqnarray}
with 
\begin{eqnarray}\label{Ubas13linrApp}
\mathcal{\hat{U}}=
\left(
\begin{array}{cc}
{\mathcal{R}^2e^{i\phi_d}-\mathcal{T}^2e^{-i\phi_d}}&
{-\mathcal{RT^*}e^{i\phi_d}-\mathcal{R^*T}e^{-i\phi_d}} \\
{\mathcal{RT^*}e^{i\phi_d}+\mathcal{R^*T}e^{-i\phi_d}}& 
{\mathcal{R^*}^2 e^{-i\phi_d}-\mathcal{T^*}^2e^{i\phi_d}}
\end{array}\right),\cr
\end{eqnarray}
where the reflection index $\mathcal{R}$ (transmission index $\mathcal{T}$) can be extracted as $\mathcal{R}=\langle\mathcal{E}_{-}(t_f)|\Psi(t_f)\rangle$ ($\mathcal{T}=\langle\mathcal{E}_{+}(t_f)|\Psi(t_f)\rangle$) when we start from  the ground state.
The eigenbasis $\{|\mathcal{E}_{-}(t)\rangle, |\mathcal{E}_{+}(t)\rangle\}$ is defined in Eqs.~\eqref{EconFHAppC} and \eqref{EconLHApp}, i.e., with requirements ($t\in[0,T/2]$)
\begin{eqnarray}
&\arg\big[\langle\mathcal{E}_{+}(t)|0\rangle\langle0|\mathcal{E}_{-}(t)\rangle\big]
=2\phi_{c},&\cr\cr
&\det\big[\hat{\mathcal{S}}(t)\big]
=\det\big[|\mathcal{E}_{-}(t)\rangle,|\mathcal{E}_{+}(t)\rangle\big]=1.&
\end{eqnarray}
For the convenience of calculation, we take a gauge transformation: 
\begin{eqnarray}
|\mathcal{E}_{-}(t)\rangle&\rightarrow& e^{-i\phi_{c}}|\mathcal{E}_{-}(t)\rangle, \cr\cr |\mathcal{E}_{+}(t)\rangle&\rightarrow& e^{i\phi_{c}}|\mathcal{E}_{+}(t)\rangle,
\end{eqnarray}
and the gauge requirements become ($t\in[0,T/2]$)
\begin{eqnarray}
&{\rm Im}\big[\langle\mathcal{E}_{+}(t)|0\rangle
\langle0|\mathcal{E}_{-}(t)\rangle\big]=0,&\cr\cr &\det\big[\hat{\mathcal{S}}(t)\big]=
\det\big[|\mathcal{E}_{-}(t)\rangle,|\mathcal{E}_{+}(t)\rangle\big]=1,& \end{eqnarray}
with associating indexes being changed as 
\begin{eqnarray}
\mathcal{R}\rightarrow e^{i\phi_{c}}\mathcal{R},~~~~ \mathcal{T}\rightarrow e^{-i\phi_{c}}\mathcal{T}.
\end{eqnarray}
As a result Eqs.~\eqref{U} and \eqref{Uelement} of the main text have been obtained.

\end{document}